\documentclass[oribibl]{llncs}

\usepackage{url}
\usepackage{amsmath}
\usepackage{amssymb}
\usepackage{graphicx}
\usepackage{tabularx}
\usepackage{booktabs}
\usepackage{eurosym}

\DeclareMathOperator{\N}{N}
\newcommand{\bsym}[1]{\boldsymbol{#1}}
\newcommand{\bchar}[1]{\textup{\textbf{\textrm{#1}}}}

\usepackage[colorlinks=true, linkcolor=blue,citecolor=blue,urlcolor=blue]{hyperref}

\usepackage{fancyhdr} 
\fancypagestyle{firststyle} {

\fancyhf{} \fancyfoot[C]{\scriptsize{Proceedings of the 2nd Multidisciplinary
    International Symposium on Disinformation in Open Online Media (MISDOOM
    2020), Leiden (online), Springer, pp.~37--51. The final publication is available at \url{https://doi.org/10.1007/978-3-030-61841-4_3}}}
}

\begin{document}

\title{A Dip Into a Deep Well: Online Political Advertisements, Valence, and European Electoral Campaigning}
\author{Jukka Ruohonen
\\\email{juanruo@utu.fi}}
\institute{Department of Future Technologies, University of Turku, Turku, Finland}

\maketitle

\begin{abstract}
Online political advertisements have become an important element in electoral campaigning throughout the world. At the same time, concepts such as disinformation and manipulation have emerged as a global concern. Although these concepts are distinct from online political ads and data-driven electoral campaigning, they tend to share a similar trait related to valence, the intrinsic attractiveness or averseness of a message. Given this background, the paper examines online political ads by using a dataset collected from Google's transparency reports. The examination is framed to the mid-2019 situation in Europe, including the European Parliament elections in particular. According to the results based on sentiment analysis of the textual ads displayed via Google's advertisement machinery, (i) most of the political ads have expressed positive sentiments, although these vary greatly between (ii)~European countries as well as across (iii) European political parties. In addition to these results, the paper contributes to the timely discussion about data-driven electoral campaigning and its relation to politics and democracy.
\end{abstract}


\section{Introduction}

\thispagestyle{firststyle} 

Political communication is increasingly affect-laden; many politicians use strong words and seek big emotions for delivering their messages. The delivery, in turn, is nowadays often done rapidly through social media and micro-targeted online advertisements. Similar delivery tactics are used to also spread outright misinformation and propaganda. There is a similarity between these and the political online communication of many politicians; both seek to appeal to emotions.

In online marketing emotions are embedded to the concept of valence. Although the origins come from psychology, there is a whole academic branch devoted to the concept in marketing research. Without delving into the details of this branch, in essence, products with a positive brand sell. By implication, it is important for marketers to try to increase positive valence expressed by consumers online. When money is involved, however, anything appearing online is subject to manipulation and exploitation. In fact, it has long been known that the online marketing industry also garners massive gray and black areas, ranging from shadowy business practices that evade existing consumer protection laws to downright criminal activities~\cite{Kshetri10}. There is also more to the concept of valence in online marketing settings. To increase the overall attractiveness of a brand, marketers often use so-called electronic word-of-mouth techniques; the goal is to spread a positive sentiment expressed by one consumer to other consumers~\cite{Ruohonen17ELMA}. There are also risks involved. For instance, customers may start to also propagate negative sentiments, leading to negative spillovers through which negative online chatter about a brand affects negatively the brand's other product segments as well as rival brands~\cite{Borah16}. In politics such risks amalgamate into strategies; to win elections, spreading negativity and misinformation may yield a good payoff.

This brief background provides the motivation for the present paper---as well as its contribution. Although there is a growing literature on online disinformation, fake news, and related topics~\cite{Pierri20, Bietti19, Wright20}, the connection of these to online marketing is seldom explicitly articulated. A further point can be made about the intermediaries through which disinformation---and marketing material---is spread. Although Facebook continues to be the platform of choice for commercial marketers, political campaigners, and miscreants alike~\cite{Bradshaw19, Dommett19}, practically the whole Web has been captured to serve advertising. Yet, much of the research has focused on social media, and, presumably due to the availability of open data, Twitter. Paid online advertisements have seldom been examined. In fact, this paper is likely the very first to explore the online political ads displayed through the advertisement gears of Google, the world's largest advertisement company.

By following recent research~\cite{Pierri20}, the paper's focus is further framed to the mid-2019 situation in Europe, including particularly the 2019 European Parliament (EP) elections in the European Union (EU). In this regard, it is worth remarking that the EU's attempts to combat disinformation and manipulation can be roughly grouped into two approaches: the General Data Protection Regulation (GDPR) on one hand and voluntary self-regulation on the other~\cite{Nenadic19}. This dual approach is a little paradoxical; while also Google has released these voluntary code-of-practice  reports for political ads~\cite{Google19c}, the company is at the same time under GDPR and other investigations by authorities in the EU and its member states. A paradox is present also in politics: many European politicians and political parties---including those who campaigned for the GDPR and who have advocated better privacy regulations in general---have been eager to market themselves online by using the tools and techniques supplied by the advertisement~industry. Divines do not always practice what they preach.\footnote{A Cyclopedia of the Best Thoughts of Charles Dickens, Compiled and Alphabetically Arranged by F.~G.~De Fontaine, New York, Hale \& Son publishers, 1872, p.~267.}

In order to present a few sensible and testable hypotheses for the forthcoming empirical exploration of textual ads displayed through Google, Section~\ref{sec: hypotheses} continues the discussion about the relation between online marketing and electoral campaigning. The dataset and the methods used are elaborated in the subsequent Section~\ref{sec: materials and methods}. Results and conclusions follow in Sections~\ref{sec: results} and \ref{sec: conclusion}, respectively.

\section{Hypotheses and Related Work}\label{sec: hypotheses}

Data-driven campaigning was one of the keywords in the 2010s politics. Throughout the decade and throughout the world, politicians and party officials were enthusiastically experimenting with new techniques for targeting electorates and influencing their opinions online~\cite{Anstead17, Jungherr16, Rafalowski19}. The tools and techniques used were exactly the same as the ones used for commercial online marketing~\cite{Chester19, Christl17}. However, things changed dramatically in the late 2010s; the 2016 presidential election in the United States and the later Cambridge Analytica scandal in 2018 were the watershed moments for the change. No longer was data-driven campaigning uncritically seen in positive light by electorates and political establishments. Manipulation, disinformation, and related concepts entered into the global political discourse. This entry was nothing unexpected from a computer science perspective; academic privacy research had pinpointed many of the risks well before these gained mainstream traction~\cite{Korolova10}. Later on, social media and technology companies sought to answer to the public uproar by traditional means of corporate social responsibility: by producing voluntary transparency reports on political ads. The reports released by Google supply the data for the present~work.

If full corporate social responsibility is taken for granted, these reports cover most of the political ads shown through the Google's vast online advertisement empire. These are paid advertisements for which a record is kept about the advertisers. Therefore, the paper's topic covers manipulation but excludes blatant disinformation, which, at least presently, unlikely occurs extensively through paid online ads. Yet, there is still a notable parallel between these ads and the genuine disinformation that is being primarily spread on---or via---social media.

Whether it is plain propaganda, indirect distractions, smear campaigns, peppering of political polarization, or suppressing participation through harassment, the tactics used tend to emphasize emotions or valence, the attractiveness or unattractiveness of a political message~\cite{Bradshaw19}. The same emotional emphasis has long been a part of online marketing~\cite{Colicev19}. Furthermore, valance provides a clear connection to political science within which negative electoral campaigning is a classical research topic. Although definitions vary, a directional definition is often used; these campaigns involve attacks against and confrontation with competing political actors~\cite{Song19}. Such campaigns have become common also in Europe through populist parties who seek to appeal to people and their emotions with criticism about establishments and the exclusion of others~\cite{Sakki17, Schmidt17}. While populism thus involves both the directional definition and the aspect of valence, there exists also an alternative definition of negative campaigning often cherished by politicians, campaigners, and consultants: because confrontations belong to politics, negative campaigning, according to the definition, is more about negative political messages that involve untruthful or deceptive claims~\cite{Walter14}. By loosely following this alternative, non-directional definition, the present work concentrates on the potential valance-rooted negativity present in online political ads.

Such negativity is neither a fully social nor an entirely political phenomenon; it contains also visible socio-technical traits. Although the so-called echo chambers would be a good example, the evidence regarding such chambers is mixed~\cite{Nguyen19}. Therefore, it is more sensible to generally assert that incivility breeds further incivility, and online platforms are not neutral actors in this breeding~\cite{Wright20}. On the technical side of this nurturing, a good example would be the 2012 experiment by Facebook to manipulate users' news feeds to determine whether emotionally positive or negative reactions could be invoked algorithmically~\cite{Christl17}. Although such proactive manipulation of masses is beyond the reach of academic research, related negativity propagation topics have been examined also in marketing \cite{Borah16} and computer science~\cite{Ozer17}. Propagation provides a powerful tool also in politics.

On the social and political side, data-driven campaigning has presumably sought to conduct many similar experiments, as testified by the Cambridge Analytica scandal. Though, the actual power and control of politicians, campaigners, and data mining companies may still be somewhat illusory; they are dependent on the existing online advertisement machinery, which, in turn, is often based on vague datasets supplied by shady data brokers, questionable machine learning, and even plain pseudo-science. Furthermore, by nature, politics are always volatile, non-deterministic, and ambivalent---by implication, it is extremely difficult to predict which particular topics become the focal topics in a given election. The 2019 EP elections are a good example in this regard: although immigration, populism, and euroskepticism were all well-anticipated topics~\cite{Pierri20}, the emergence of climate change as a topic was hardly well-predicted. The results from these European elections also polarized around these topics; populist euroskeptic parties won, but so did pro-Europe and green parties. Given this background, the first hypothesis examined in the forthcoming empirical analysis can be stated~as:
\begin{description}
\item[\normalfont{H}$_1$]{\textit{Reflecting the current political polarization and the particular themes in the 2019 EP elections, the online political ads that were shown in Europe around mid-2019 tended to exhibit negative sentiments and negativity in general.}}\label{h: negativity}
\end{description}

The literature on negative campaigning allows to refine this Hypothesis $\textmd{H}_1$ into a couple of additional, inferential hypotheses. In particular, it has been observed that party systems and characteristics of political systems in general  affect negative campaigning and its prevalence~\cite{Elmelund13, Walter14}. In essence, two-party systems have often been seen as more prone to negative campaigning than the multi-party systems and coalition governments that are typical to most European countries. Therefore, it seems justified to also posit the following hypothesis:
\begin{description}
\item[\normalfont{H}$_2$]{\textit{The sentiments---whether positive or negative---expressed in the political online ads around the 2019 EP elections varied across the EU member states.}}\label{h: countries}
\end{description}

A corollary Hypothesis~$\textmd{H}_3$ logically follows:

\begin{description}
\item[\normalfont{H}$_3$]{\textit{The sentiments expressed in the mid-2019 European online political ads varied not only across the EU member states but also across political parties.}}\label{h: parties}
\end{description}

As party systems vary across Europe, so do parties, contextual factors, campaigning strategies, and political cultures. Besides this truism, Hypothesis~$\textmd{H}_3$ can be justified with existing observations that different parties tend to use online campaigning techniques differently~\cite{Anstead17, Jungherr16,Rafalowski19}. Finally, it should be noted that neither $\textmd{H}_2$ nor $\textmd{H}_3$ are logically dependent on the answer for Hypothesis~$\textmd{H}_1$.

\section{Materials and Methods}\label{sec: materials and methods}

\subsection{Data}

\begin{figure}[th!b]
\centering
\includegraphics[width=\linewidth, height=4cm]{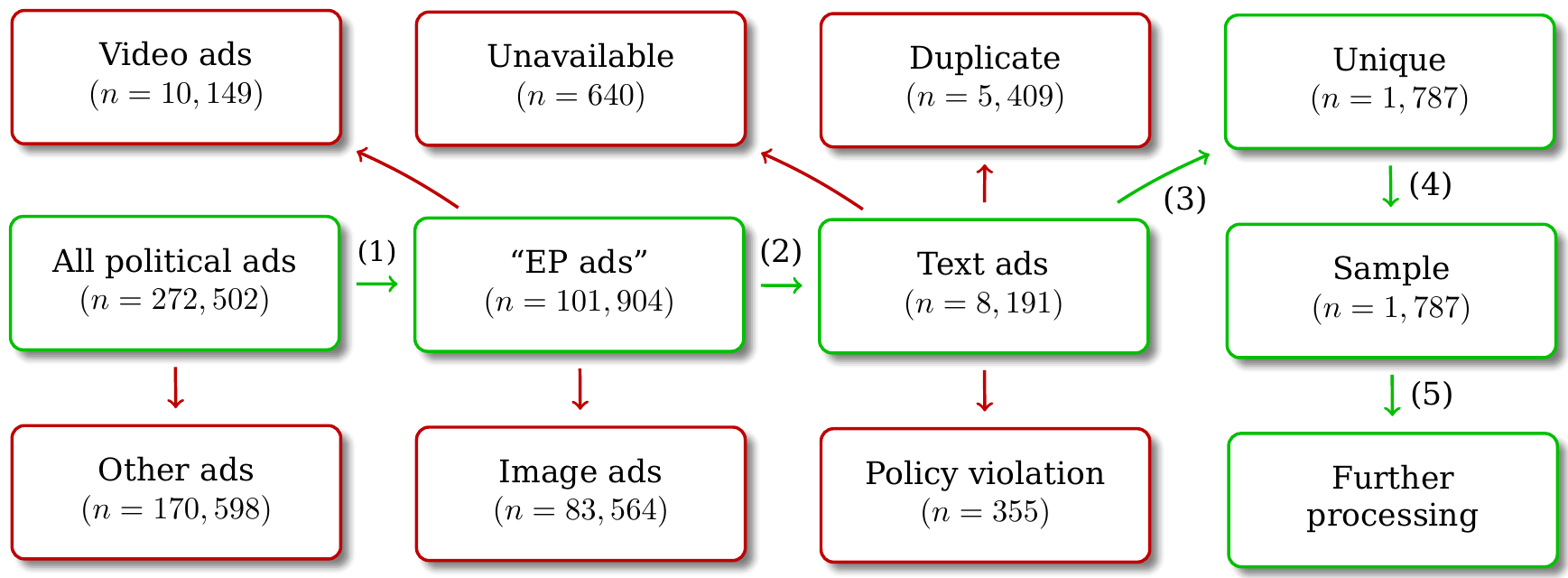}
\caption{The Construction of the Sample }
\label{fig: preprocessing}
\end{figure}

The dataset is based on Google's \cite{Google19a} transparency reporting on the political advertising in the European Union. The following seven important points should be enumerated about the dataset and its pre-processing for obtaining the sample:

\begin{enumerate}
\itemsep 2pt
\item{The EU itself is only used by Google to distinguish the geographic origins of the authors of the political ads. By implication, the data does not separate advertisements exclusively \textit{about} the EU and its elections---nor does it distinguish advertisements potentially placed \textit{by} the EU and its institutions. However, information is available about elections targeted by an advertiser. Given this information, the sampling of observations was restricted to those advertisers who had announced having advertised in the 2019 EP elections.}
\item{Only textual advertisements were included in the sample. As can be seen from Fig.~\ref{fig: preprocessing}, most of the political ads placed through Google were in fact videos and images. The textual advertisements are those typically seen as so-called paid banners in the company's search engine results, while the political video advertisements typically appear in YouTube, and so forth.}
\item{All textual advertisements in the sample were further translated to English by using Google's online translation engine. By and large, this automatic translation is necessary because contemporary text mining frameworks remain limited in their coverage of the multiple languages spoken in Europe.}
\item{Duplicate textual advertisements were excluded. This exclusion was done with simple string matching before and after the translation: if two ads contained the exact same text, only one of these was included in the sample.}
\item{Given the lexicon-based sentiment analysis techniques soon described, only minimal pre-processing was applied to the translated ads. Namely: the strings ``\texttt{no.}'', ``\texttt{No.}'', and ``\texttt{NO.}'' were excluded because the sentiment techniques tend to equate these to negations, although in the present context these refer to campaigning with a candidate's number in a particular election.}
\item{No data was available for some advertisements due to third-party hosting of the advertisements and violations of Google's policies~\cite{Google19c} for political ads. Given the ongoing debate about online political ads in general, the quite a few policy violations are particularly interesting, but, unfortunately, no details are provided by Google regarding the reasons behind these violations.}
\item{The data is very limited and coarse with respect to targeting and profiling~\cite{Chester19}.}
\end{enumerate}

The last point requires a brief further comment. Although hosting and technical traceability have recently been under regulatory scrutiny~\cite{Haenschen19}, the micro-targeting, mass-profiling, and manipulation aspects have received most of the general political attention~\cite{Chester19, Christl17, Dommett19}. In this respect, Google seems to have aligned itself more toward Facebook than toward Twitter and Spotify, both of which have banned all political ads in their platforms. In fact, a spokesperson from Google recently assured that the company has ``\textit{never offered granular microtargeting of election ads}'', but, nevertheless, since the beginning of 2020, it now only allows targeting of political advertisements according to age, gender, and postal code~\cite{Google19b}. Some data about age and gender targeting is also available in the transparency reports. In theory, this data could be useful for continuing the work on Google's demographic profiling~\cite{Tschantz18}, but, in practice, the data is of little practical use. For instance: from all advertisement campaigns in the raw dataset ($n = 46,880$), which group multiple ads, about 80\% have not specified gender-based targeting. The second largest group (18\%) is something labeled as ``male, female, unknown gender'', which, more than anything, foretells about (perhaps intentional) construct validity problems affecting the transparency reporting.

An additional point should be made about the longitudinal scope of the sample. The sample covers a period of about five months. The earliest and latest advertisements in the dataset are from 20 May 2019 and 6 October 2019, respectively. The starting date is constrained by data availability; in general, Google does not provide earlier data. The ending date, in turn, is framed with the date of obtaining the raw dataset (9~October 2019). Given the varying lengths of electoral campaigns, the 2019 European Parliament elections (23--26 May) are thus only partially covered. Even though the coverage captures only the few late days in the campaigning for the EP elections, it seems fair to assume that these were also the dates of particularly intense campaigning. The point is important especially in the online context, which does not require lengthy upfront planning. In other words, online political ads are easy to place even for last-minute probes.

However, even with the noted restriction of the sample to those advertisers who had advertised in the EP elections, also other elections and referendums are potentially covered because these advertisers may have advertised also in other occasions. Given the longitudinal scope, these occasions include: the Irish referendum on divorce (24 May) and the Romanian referendum on corruption (26 May), the federal election in Belgium (26 May), the second round in the Lithuanian presidential election (26 May), the Danish and Greek parliamentary elections (5 June and 7 July, respectively), the lower house election in Austria (29 September), and the Portuguese parliamentary election (6 October). In addition, the Brexit saga is visible also in the sample analyzed. Furthermore, politicians, party officials, interest groups, and individuals may also place online ads for general advocacy and publicity reasons without a clear electoral target~\cite{Dommett19}. All this said, qualitative observations and a few keyword-based searches indicate that many of the ads sampled explicitly or implicitly refer to the 2019 EP elections.

\subsection{Methods}\label{subsec: methods}

Sentiment analysis refers to a group of computational methods to identify subjective information and affective states. In the text mining context these methods can be roughly grouped into machine learning and lexicon-based approaches. Two simple lexicon-based methods are used in the present work: the algorithms of Liu et al.~\cite{Liu05} and Nielsen~\cite{Nielsen11}, as implemented in an R package~\cite{syuzhet}. Both rank the sentiment of a document according to the number of times manually labeled negative and positive words appear in the document. In addition, the slightly more sophisticated method of Hutto and Gilbert~\cite{VADER} is used, as implemented in a Python package~\cite{NLTK19}. This method augments the lexicon-based approach with a few (deterministic) rules on the grammar and style used in a document. Based on a subjective evaluation, no normalization is used for the lexicon-based algorithms, while the algorithm of Hutto and Gilbert is scaled to the unit interval. All three methods are tailored for text mining of social media data. Therefore, the methods seem also suitable for analyzing the textual political advertisements delivered through Google. Akin to messages in Twitter, these ads are short and up to a point; the mean character count of the sample is only $118$ characters.

The first Hypothesis~$\textmd{H}_1$ is examined with descriptive statistics. Regression analysis is used for examining $\textmd{H}_2$ and $\textmd{H}_3$. To examine the two hypotheses in a single linear model, the following random effects (or multi-level) model is used:
\begin{equation}
y_{ij} = \alpha_j + \bchar{x}^\prime_{ij}\bsym{\beta} + \varepsilon_{ij} ,
\quad~i = 1, \ldots, 1787,
\quad~j = 1, \ldots, 63 ,
\end{equation}
where $y_{ij}$ is a sentiment score from a given algorithm for the $i$:th ad, $\alpha_j$ is a random effect for the $j$:th political party, $\bchar{x}^\prime_{ij}$ is a row vector of non-random (fixed) independent variables, $\bsym{\beta}$ is a regression coefficient vector, $\varepsilon_{ij}$ is the error,
\begin{equation}\label{eq: random effects}
\alpha_j \sim \N(0, \sigma^2_\alpha)
\quad\textmd{and}\quad
\varepsilon_{ij} \sim \N(0, \sigma^2_\varepsilon) ,
\end{equation}
where $\N(\cdot)$ denotes the normal distribution with a mean of zero and variance~$\sigma^2$.

Thus, the effects for the political parties are treated as random variables, which are both mutually independent and independent from the errors $\varepsilon_{ij}$. In contrast, the country effects are embedded to $\bchar{x}^\prime_{ij}$ together with other control variables enumerated in Table~\ref{tab: independent variables}. This model is generally necessary because the country and party effects cannot be both included as normal (fixed) independent variables due to multicollinearity. Furthermore, the countries cannot be easily modeled as random effects because some ads have been shown in many~countries.

Although further variability could be examined by allowing also the slope coefficients in $\bsym{\beta}$ to vary across the parties, the model specified is enough to answer to the two hypotheses. It is fitted for the results from all three sentiment algorithms using well-known and well-documented R packages~\cite{Baayen08, lme4, lmerTest}. If the answers to both $\textmd{H}_2$~and~$\textmd{H}_3$ are positive, in essence, $\hat{\sigma}^2_\alpha \not\approx 0$ and some of the estimated coefficients in $\hat{\bsym{\beta}}$ for the countries are non-zero and statistically significant. In addition, the packages used provide a function to test the random effect terms $\hat{\alpha}_j$ with a likelihood ratio test. Likewise, Akaike's information criterion (AIC) is used to compare the full (unrestricted) model to a restricted model without the country effects. The simple random effects model and the basic statistical checks outlined are sufficient because the goal is not prediction, which might entail model validation with bootstrapping, cross-validation, or related~techniques.

\begin{table}[th!b]
\centering
\caption{Independent (Fixed) Variables}
\label{tab: independent variables}
\begin{tabularx}{\linewidth}{clX}
\toprule
Mnemonic && Description \\
\hline
DAYS && A continuous variable measuring the number of days an ad was shown. \\
\cmidrule{3-3}
IMPR && Three dummy variables for the number of Google-defined ``impressions'' an ad got; the reference variable is less than ten thousand impressions. \\
\cmidrule{3-3}
EURO && Three dummy variables for the upper bound of the cost of an ad; the reference variable is $50$\euro~(the maximum dummy variable denotes~$60,000$\euro). \\
\cmidrule{3-3}
AGET && A dummy variable that takes the value one in case any of the campaigns to which an ad belonged had specified any kind of age-based targeting. \\
\cmidrule{3-3}
GENT && Defined analogously to AGET, but for gender-based targeting. \\
\cmidrule{3-3}
MULT && A dummy variable scoring $1$ if an ad was displayed in multiple countries. \\
\cmidrule{3-3}
CNTR && Twenty-five dummy variables for the countries in which an ad was shown. \\
\bottomrule
\end{tabularx}
\end{table}

A further point should be made about the identification of political parties. This identification was done manually. For unclear cases, open source intelligence (a.k.a.~Google and Wikipedia) was used to check whether the name of an advertiser referred to an European political party. On the one hand, the mapping includes cases whereby a local or a regional chapter of a clearly identifiable party had placed the given political ad; on the other, electoral alliances had to be excluded from the identification.  Although about 72\% of all political ads could be mapped to parties, it should be emphasized that many of the political ads were placed by different support associations, marketing companies, and even individuals on behalf of some particular politicians and candidates. National election laws also differ between the EU member states with respect to the general rules on electoral campaigning. Currently, only eleven member states have specific legislations in place regarding mandatory transparency of online political ads~\cite{Nenadic19}. Needless to say, these judicial aspects are an important element in the debate about political online ads---and the sample also contains some cases in which a vague support association in one country had advertised in another country.

\section{Results}\label{sec: results}

\begin{figure}[th!b]
\centering
\includegraphics[width=\linewidth, height=4cm]{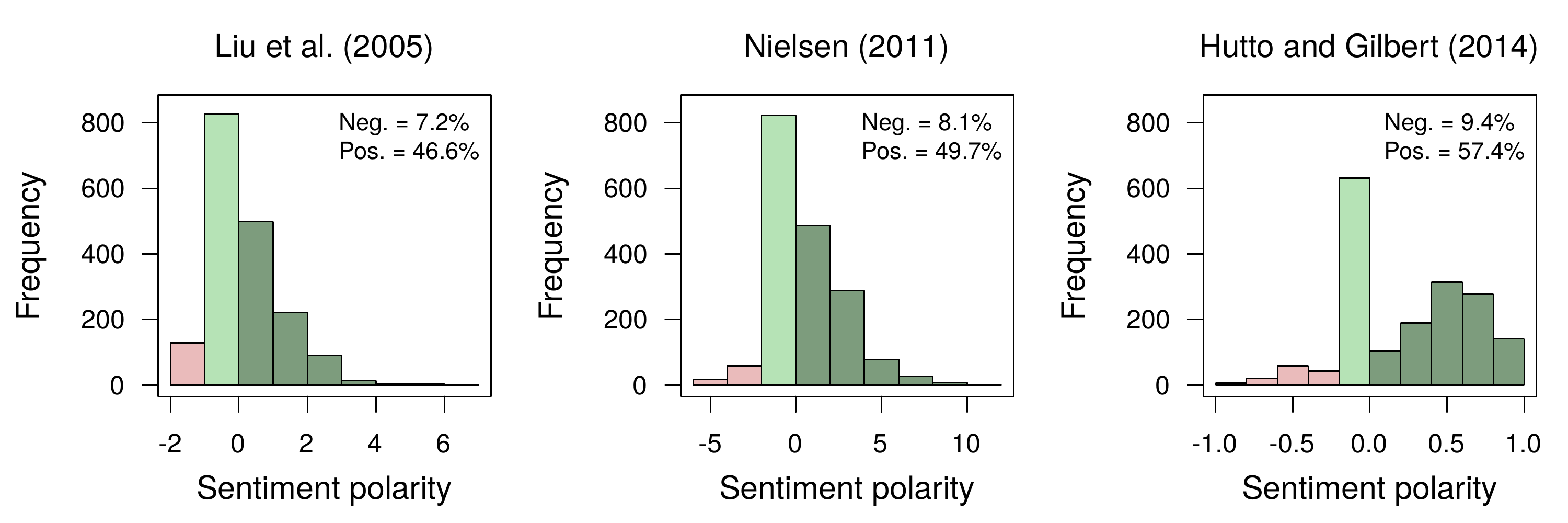}
\caption{Sentiment Polarity According to Three Algorithms}
\label{fig: sentiments}
%
\vspace{15pt}
%
\centering
\includegraphics[width=\linewidth, height=8cm]{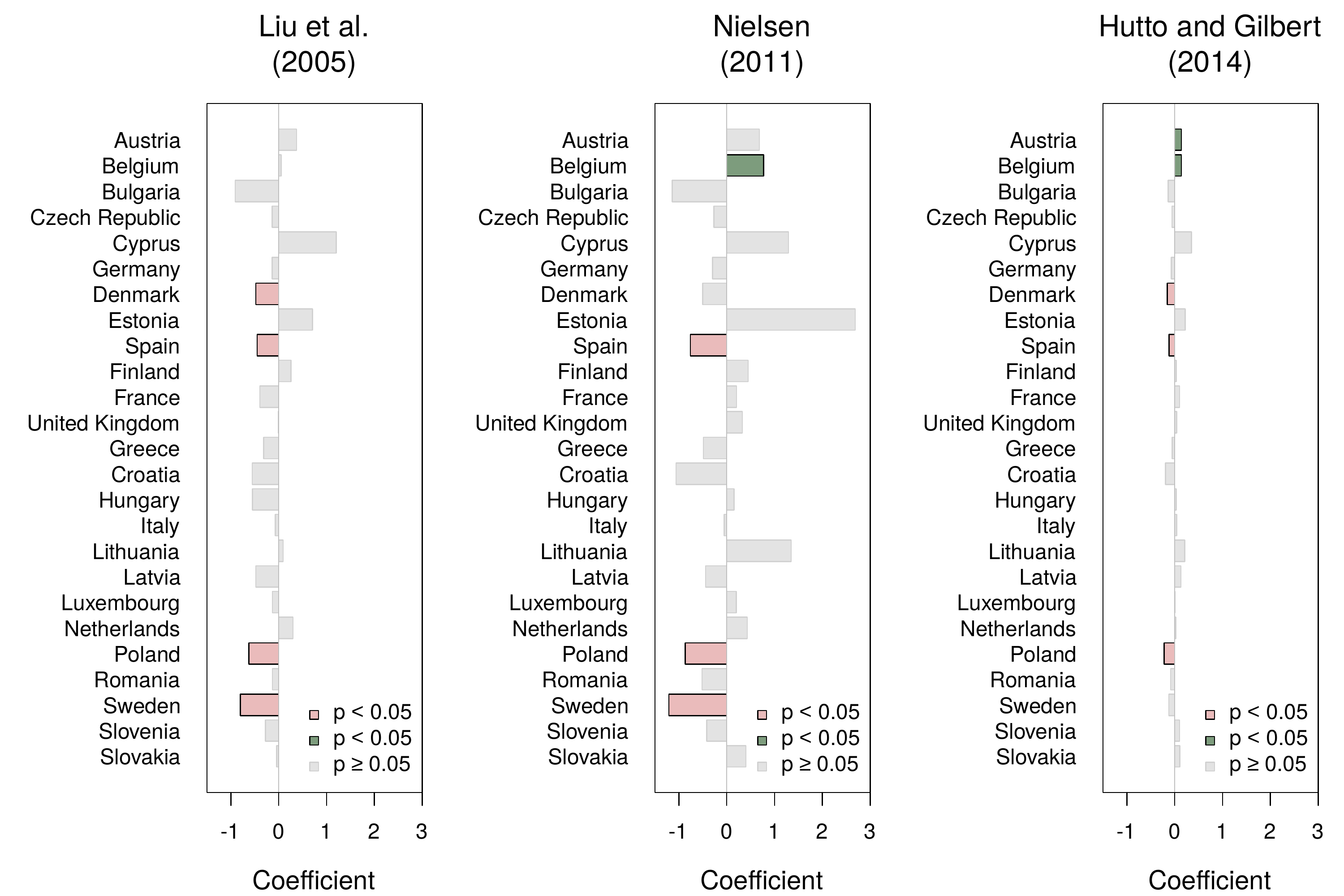}
\caption{Country Effects (regression coefficients)}
\label{fig: countries}
\end{figure}

All three sentiment algorithms indicate pronouncedly non-negative valence. As can be seen from Fig.~\ref{fig: sentiments}, only less than 10\% of the ads have a negative sentiment polarity according. Depending on an algorithm, zero-valued sentiments account for about 33--46\% of all political ads in the sample. For the two lexicon-based algorithms, these ``neutral'' (zero-valued) scores imply that no word in an ad was tagged as positive or negative, or that an equal number of words were tagged as positive and negative. Thus, the dataset and the algorithms do not support~$\textmd{H}_1$.

\begin{table}[th!b]
\centering
\caption{AIC Values from Unrestricted and Restricted Models}
\label{tab: aics}
\begin{tabular}{lcccccc}
\toprule
\qquad\qquad& Liu et al.~(2005) &\quad& Nielsen (2011) &\quad& Hutto and Gilbert (2014) \\
\cmidrule{2-6}
Restricted
& $5316$ && $7570$ && $1389$ \\
Unrestricted
& $5271$ && $7520$ && $1326$ \\
\bottomrule
\end{tabular}
\end{table}

\begin{table}[th!b]
\centering
\caption{Variances of the Random Effects}
\label{tab: variance}
\begin{tabular}{lcccccc}
\toprule
\qquad\qquad& Liu et al.~(2005) &\quad& Nielsen (2011) &\quad& Hutto and Gilbert (2014) \\
\cmidrule{2-6}
$\hat{\sigma}^2_\alpha$
&  $0.168$ && $0.697$ && $0.035$ \\
$\hat{\sigma}^2_\varepsilon$
& $1.054$ && $3.698$ && $0.114$ \\
\bottomrule
\end{tabular}
\end{table}

However, there is some variation across the EU countries in which the online ads were shown. This observation can be seen from Fig.~\ref{fig: countries}, which shows the estimated regression coefficients for the country effects. Possibly due to the lack of normalization, there is interesting variation across the three algorithms; in particular, the coefficients are much smaller in magnitude for the sentiments computed with the algorithm of Hutto and Gilbert. Only the negative coefficients for Poland and Spain are statistically significant for all three algorithms. While the statistically significant coefficients are not substantial in magnitude for any of the algorithms, the AIC values in Table~\ref{tab: aics} still indicate small improvements. In other words, the  country effects are worth retaining; there is weak support for~$\textmd{H}_2$. As for the other control variables in Table~\ref{tab: independent variables}, only one of the dummy variables for EURO is significant across all three algorithms. As seen from Table~\ref{tab: variance}, the variances of the random party effects are also non-zero. The \texttt{ranova} function~\cite{lmerTest} further indicates statistical significance of the random effects for all three sentiment algorithms. Thus, also Hypothesis~$\textmd{H}_3$ can be accepted.

\begin{figure}[p!]
\centering
\includegraphics[width=\linewidth, height=18.5cm]{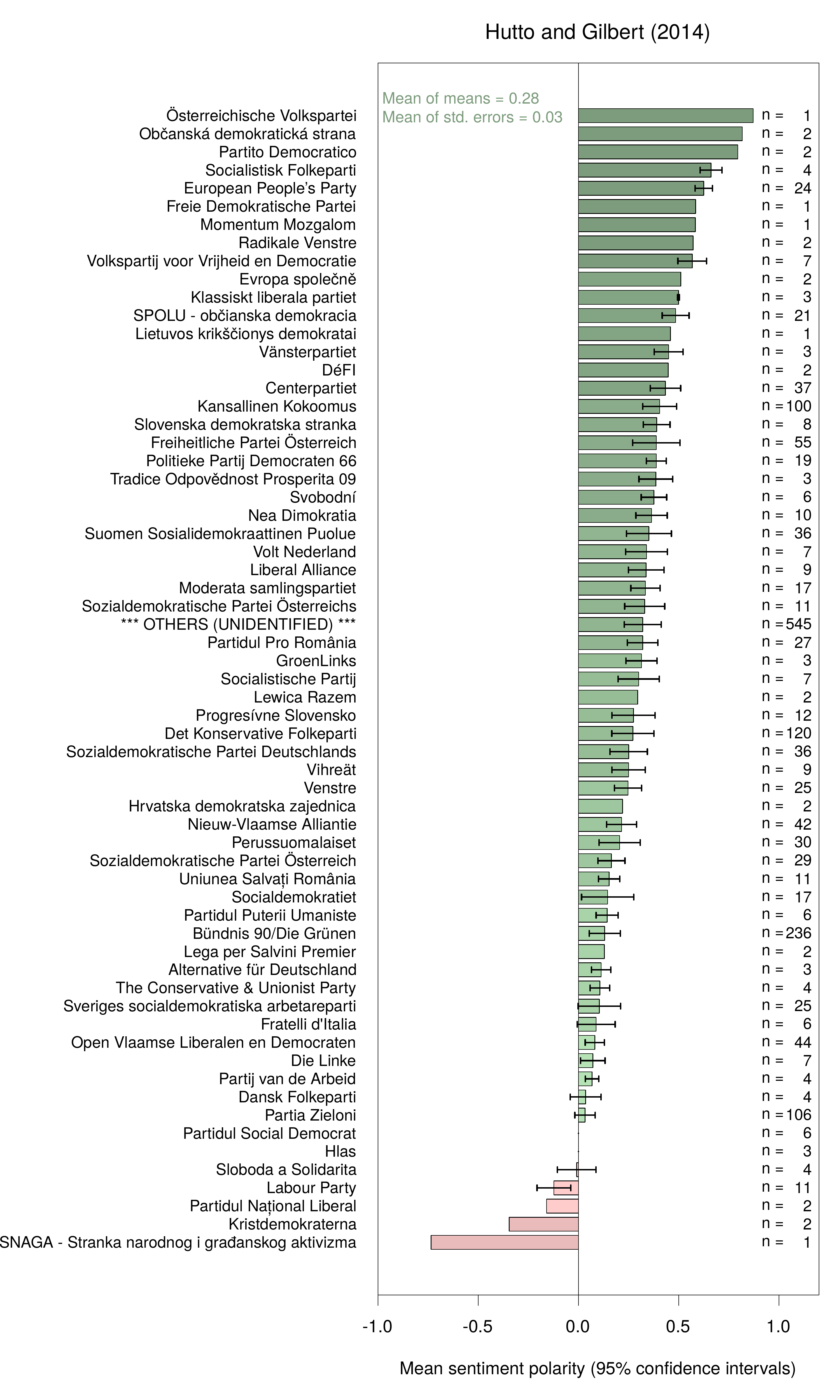}
\caption{Average Sentiment Polarity Across Political Parties}
\label{fig: parties}
\end{figure}

Indeed, a substantial variation exists both in terms of the ads placed and the sentiments expressed by the manually identified political parties (see Fig.~\ref{fig: parties}). However, it is difficult to say anything specific about the potential explanations behind this variation. For instance, many of the euroskeptic parties---including Alternative f\"ur Deutschland (Germany), Dansk Folkeparti (Denmark), Sloboda a Solidarita (Slovakia), or Fratelli d'Italia and Salvini's Lega in Italy---rank clearly below the average sentiment polarity scores. While this observation is expected, some other euroskeptic parties, such as Freiheitliche Partei \"Osterreich (Austria) and Svobodn\'i (Czech Republic), have placed ads with clearly positive sentiments. On average, these ads even expressed more positive sentiments than those seen in the ads of B\"undnis 90/Die Gr\"unen (the green party in Germany), for instance.

In general, the variability traces to the particular themes in the EP elections and the national styles of political communication used in the online ads placed through Google's kingdom. For instance, one German ad started with an indirect rhetorical question about ``\textit{whether nationalists, right-wing populists and right-wing radicals destroy Europe}'', and continued with an indirect answer: ``\textit{or whether Europe remains a place of freedom, peace and cohesion}''. Another ad likewise ended to a slogan: ``\textit{for courage, cohesion and humanity instead of fear, hatred and exclusion}''. Both are good examples about a political advertisement style through which negative and positive sentiments balance each other out. A further explanation relates to the climate change that was a pronounced theme particularly in Germany. This theme was accompanied with many ads using contentious words with a negative tone, such as crisis, fight, suffer, failure, or ``\textit{a healthy agriculture without poison and animal cruelty}''. Rather similar national explanations apply to Poland and Croatia, the two countries with the lowest average sentiment polarity scores. With respect to Poland, the explanation has nothing to do with euroskepticism; instead, there were a few particular candidates who campaigned online with slogans such as ``\textit{more illegal dumps and smog over Silesia}'', ``\textit{scandal needs clarification}'', ``\textit{fight low emissions}'', and so forth.  Such slogans reflect the online campaigning strategies of Partia Zieloni, the Green Party~\cite{Rafalowski19}. These brief qualitative examples reinforce the positive answers to Hypotheses~$\textmd{H}_2$ and $\textmd{H}_3$. To slightly correct the wording used to postulate these hypotheses, it seems fair to conclude that the sentiments expressed in the online ads vary simultaneously both across and within the EU member states.

\section{Conclusion}\label{sec: conclusion}

This exploratory paper examined the timely topic of online political advertisements. By using a dataset of textual ads displayed through Google's online advertisement machinery and focusing on the mid-2019 situation in Europe, including the EP elections in particular, three hypotheses were presented for the exploration with sentiment analysis. The first one ($\textmd{H}_1$) was framed with negativity---a distinct trait of negative electoral campaigning as well as a factor in valence-based online marketing in general. This hypothesis is not supported by the dataset: most of the online political ads shown in Europe have exhibited neutral or positive sentiments. Although the simple regression estimation strategy conducted does not allow to explicitly compare $\textmd{H}_2$ against $\textmd{H}_3$, it seems sensible to conclude that while there exists variation across the European countries observed, variation is also present with respect to political parties and their local or regional chapters. Further variation is presumably present in terms of particular advertisers, whether party officials, associations, consultants, marketing companies, or individual citizens placing ads on behalf of parties or candidates.

Three limitations can be noted. First, the machine-translation used likely causes inaccuracies---after all, a language's small nuances are often important particularly in political communication. The second limitation directly follows: only three simple sentiment algorithms were examined, and all of these were limited to English. Third, the empirical exploration was explicitly limited to textual ads, which, however, constitute only a minority of the political ads placed through Google's platforms (see Fig.~\ref{fig: preprocessing}). Patching these limitations offer good opportunities for further work in computer science. While multi-language sentiment algorithms are generally needed, so are specific lexicons tailored for online political communication. However, a satisfactory solution likely necessitates collaboration between computer and political scientists. For both scientists, a whole new realm also opens with a question of how to analyze disinformation, political manipulation, and sentiments expressed in these with image and video datasets.

But there are also many questions to which computers cannot answer---and with which computers should not be perhaps allowed to even interfere.  Democracy and politics are among these. While the transparency of algorithms is often touted as a path forward~\cite{Kirkpatrick20}, many of the problems are located deeper within the platforms. Thus, even with the limitations discussed, there are some lessons to be learned in this regard. Although $\textmd{H}_1$ was rejected and neutral sentiments have been common, all three algorithms still indicate a large amount of positive sentiments in the political ads. This observation can be used to argue that valence-based campaigning is widely practiced. Like with online marketing, such campaigning is partially explained by the technical constraints imposed by the advertising platforms. Short taglines with catchy sentimental words---whether positive or negative---are also what the platforms are imposing upon campaigners and political advertisers. As a consequence, the room for argumentation, discussion, debate, and ``evidence-based politics'' arguably shrinks even further.

A final point can be made about regulation. In the EU elections are regulated by national laws, and there are no cues that the EU itself would be willing to intervene. At the same time, according to a recent voluntary transparency report~\cite{Google19c}, Google detected $16,690$ EU-based accounts that violated the company's misrepresentation policies between the first of May 2019 and 26 May 2019. The sample examined aligns with this number; about 12\% of the EP-related textual ads were unavailable either due to policy violations or due to third-party hosting. These numbers hint that also Google has a problem with its self-regulation of political ads. But politics are always about power, and platforms provide one way to achieve and maintain power. In other words, regulating online political ads is difficult not only because of rights, freedoms, and legal hurdles~\cite{Bietti19}, but also because divines do not always practice what they~preach.

\subsection*{Acknowledgements}

This research was funded by the Strategic Research Council at the Academy of Finland (grant no.~327391).

\bibliographystyle{splncs03}

\end{document}